\documentclass[11pt]{article}

\usepackage[dvipdfmx]{graphicx}
\usepackage{url}
\usepackage{epic}
\usepackage{eepic}
\usepackage{fancybox}
\usepackage{here}
\usepackage{soul}

\begin{document}

\title{Resource allocation method using tug-of-war-based synchronization}

\author{
Song-Ju Kim${}^{\dag}$${}^{\ddag}$${}^{\ast}$, Hiroyuki Yasuda${}^{\S}$, Ryoma Kitagawa${}^{\P}$, and Mikio Hasegawa${}^{\P}$\\
\\

${}^{\dag}$ SOBIN Institute LLC, Kawanishi, Japan\\
https://sobin.org\\
${}^{\ddag}$ Graduate School of Media and Governance, Keio University, Fujisawa, Japan\\
${}^{\S}$ Graduate Schools for Law and Politics, The University of Tokyo, Tokyo, Japan \\
${}^{\P}$ Department of Electrical Engineering, Tokyo University of Science, Tokyo, Japan\\
${}^{\ast}$Email: kim@sobin.org
}

\maketitle

\abstract
{\bf 
We propose a simple channel-allocation method based on tug-of-war (TOW) dynamics, combined with the time scheduling based on nonlinear oscillator synchronization to efficiently use of the space (channel) and time resources in wireless communications.
This study demonstrates that synchronization groups, where each node selects a different channel, are non-uniformly distributed in phase space such that every distance between groups is larger than the area of influence.
New type of self-organized spatiotemporal patterns can be formed for resource allocation according to channel rewards. 
}
\endabstract

%\begin{keyword}
\noindent
Keyword:\\
Resource allocation, channel allocation, collision avoidance, reinforcement learning, multi-armed bandit problem, synchronization
%\end{keyword}

\newpage

\section{Introduction}

Following the COVID-19 pandemic, humanity is increasingly demanding a more efficient and equitable use of resources~\cite{fair}.
One of the key issues is the allocation of communication resources.
The demand for the Internet of Things (IoT) applications has recently been rapidly increased remarkably. 
The use of long-distance and power-saving communications, particularly low-power wide-area networks (LPWA), is expected in IoT communication.
However, due to its characteristics, frequency congestion has become a serious problem, necessitating the use of a dynamic multi-channel selection method.

Using lightweight reinforcement-learning, called ``tug-of-war (TOW) dynamics~\cite{tow,tow2}," Ma et al. proposed an autonomous channel selection method in our previous study~\cite{ma}.
All nodes learned the channel status by acknowledgement (ACK) in a repeated cycle of sleep, wake, data transmission, and sleep.
We also implemented this algorithm in $IEEE802.15.4e/4g$-compliant IoT devices and experimented with  $50$ IoT devices dynamically selecting five channels. 
However, this method did not address time-scheduling optimization for overall node transmission timings.

For efficient use of space (channel) and time resources, we present the first model toward achieving ``an economic allocation of resources."
In this study, we propose a simple channel-allocation method based on TOW dynamics, combined with the time scheduling based on the synchronization of nonlinear oscillators.
The synchronization is named ``TOW-based synchronization" because of the competing pushes and pulls that occur in phase space.
In other words, the TOW behaviors are active in both space and time in our new model.

\section{Model}

We have made the following assumptions in constructing our new model.
\begin{enumerate}
\item
There are $M$ nodes and $N$ communication channels.
\item
Regardless of whether there is any interaction between the nodes, 
each node's phase $\theta_i (t)$ ($i$$=$$1$, $\cdots$, $M$) is updated by adding $\Omega$ at every discrete time step.
\item
Each node can transmit a packet at every discrete time step through the channel selected via TOW dynamics~\cite{tow}.
\item
Each packet transmission does not affect any other transmissions only if those phase distances are larger than the area of influence $\phi_{th}$.   
In this study, we fix $\Omega = $$\phi_{th}$ for all nodes. 
\item
At some time steps (e.g. $\theta _i$$=$$0$), each node not only advances by angle $\Omega$ but also interacts with other nodes by the interaction term 
$\pm K \sin(\phi)$ which represents ``push $(-)$" or ``pull $(+)$," where $K$ represents the coupling strength (parameter), and $\phi$ represents the phase difference.
\item
Each node receives a reward ($+1$ or $-\omega$) according to the success (winning at the slot machine of selected channel (say $\# k$), whose reward probability $P_k$ under the condition that there is no node using the same channel within  $\phi_{th}$ ) or failure (otherwise). 
\end{enumerate}

\newpage

In this study, we propose the following channel-allocation method using the TOW-based synchronization which is based on the Kuramoto's de-synchronization model~\cite{kuramoto, desynch};  
For node $i$ $=$$1, \cdots, M$, each phase $\theta_i (t)$ is updated by the following equation:
\begin{eqnarray}
  \theta_i (t+1) & = & \theta_i (t) + \Omega - K \sum_{| \theta_j - \theta_i | < \phi_{th}}^{} (2 \delta_{s_j(t), s_i(t)} -1) \sin(\theta_j (t) - \theta_i (t)),   \label{eq:model} \\
 s_i (t) & = & argmax_{k} (X_k(t)), \hspace{1mm} for \hspace{1mm} k=1, \cdots, N.
 \label{eq:model2}
\end{eqnarray}
Here, $\delta_{i, j}$ is the Kronecker-Delta.

The selected channel $s_i (t)$ for node $i$ was determined by the following TOW dynamics of $X_k (t)$  ($k = 1, \cdots, N$)~\cite{tow};  

\begin{eqnarray}
  X_k (t+1) & = & Q_k (t+1) - \frac{1}{N-1} \sum_{l=1, l \neq k}^{N} Q_l (t+1) + \xi(t+1),   \label{eq:tow} \\
  Q_k (t+1) & = & R_k (t) + \alpha Q_k (t) , \label{eq:tow2} \\
  R_k (t) & = & \left\{ 
  \begin{array}{rr}
  +1  & (success),\\
  -\omega &  (failure).  \label{eq:tow2} 
\end{array} \right.   
\end{eqnarray}
Here, $\xi(t)$ is an arbitrary random noise, $\alpha$ is the memory parameter, and $R_k (t)$ is the reward from each slot machine whose reward probability is $P_k$. 

Figure 1 shows the interaction term between nodes for each phase difference.
The phase difference between two nodes increases if the same channel is selected by both nodes (de-synchronization), whereas the phase difference decreases if different channels are selected (synchronization).
This TOW-like ``push and pull dynamics" between nodes in phase space can organize grouping (synchronization group) of various selection nodes.
The groups can be distributed in the phase space such that every distance between groups should be larger than $\phi_{th}$.

\begin{figure}[htb]
\begin{center}
\includegraphics[height=50mm]{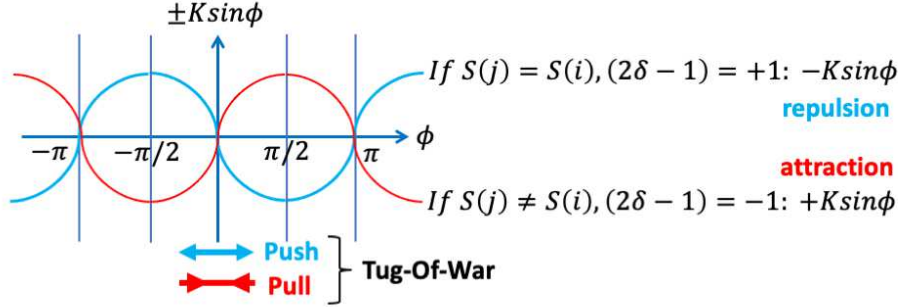}
\end{center}
\caption{The interaction term between nodes for each phase difference.}  
\label{fig:1}
\end{figure}

\subsection{Results}

In this study, we fix $\Omega = $$\phi_{th}$$= \pi /4$ for all nodes, and only once when crossing the $\theta_i$ $= 0$ (once in eight time steps), the interaction term $\pm K \sin(\phi)$ works.
We confirmed that self-organized spatiotemporal patterns can be formed.
Synchronization groups, particularly those where each node selects a different channel, are non-uniformly distributed in phase space such that every distance between groups is larger than $\phi_{th}$ (area of influence), as expected.

Figure 2(a) shows an example of our computer simulation results.
We used the following parameters: $M=10$, $N=5$, $\phi_{th}$$=$$\pi/4$, $\alpha$$=$$0.95$, $Amp$$=$$0.1$ (noise amplitude), $K$$=$$0.5$, $P_1$$=$$0.1$, $P_2$$=$$0.2$, $P_3$$=$$0.3$, $P_4$$=$$0.4$, and $P_5$$=$$0.5$ ($\omega$ is calculated by the system itself.).
Four synchronization groups can be formed in phase space.
Each red number denotes the node number which selected the channel $\#5$ (highest reward probability).

Figure 2(b) shows a typical example for two time-series of the phase of node $\#1$ and $\#7$.
The phase distance is fixed  ($203$$^{\circ}$) after the emergence of the self-organizing pattern.
In Fig. 2(b), each distance between diagonal stripes is $45$$^{\circ}$ because we set $\phi_{th}=\pi / 4$.
The difference $203$$^{\circ}$ can be seen as $23$$^{\circ}$ because $203 - (45$ $\times$ $4) $$=$$23$ in the figure.  

\begin{figure}[htb]
\begin{center}
\includegraphics[height=50mm]{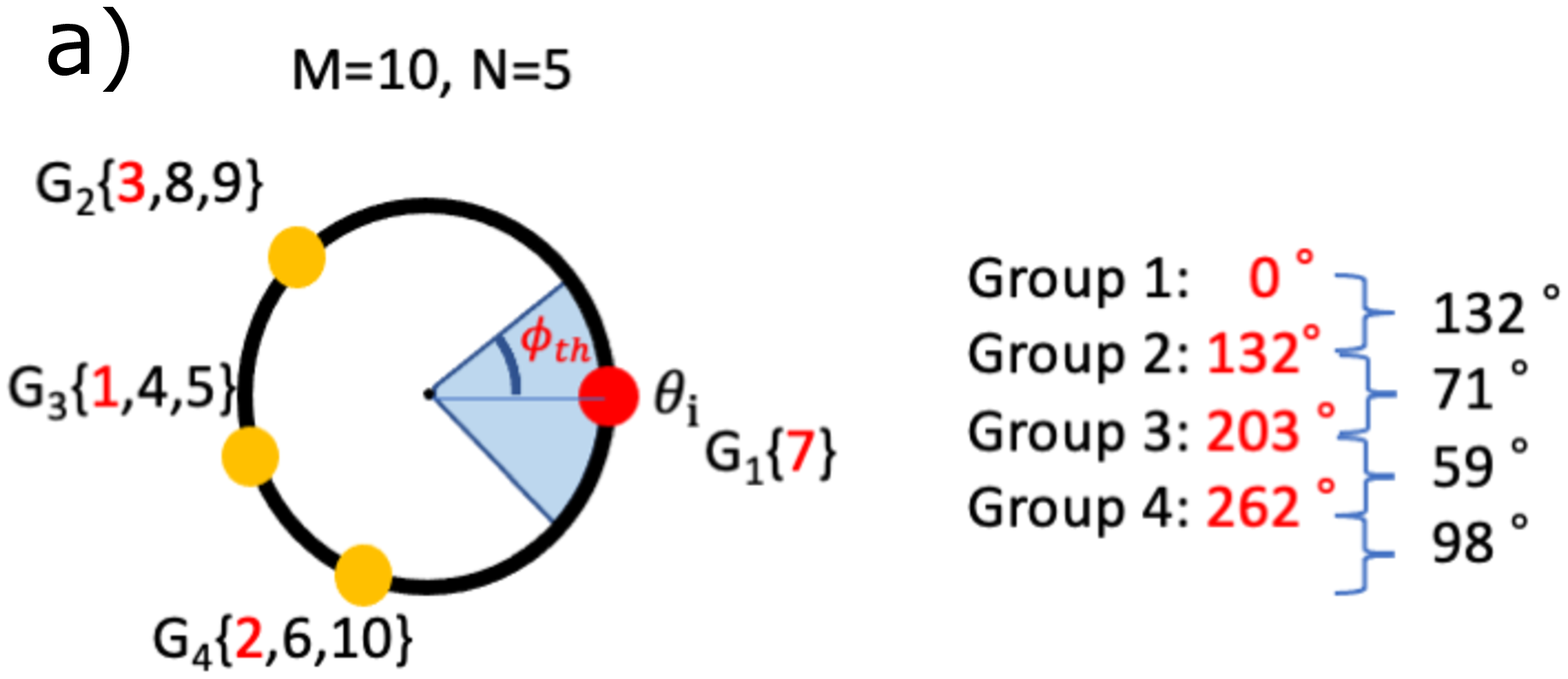}
\includegraphics[height=80mm]{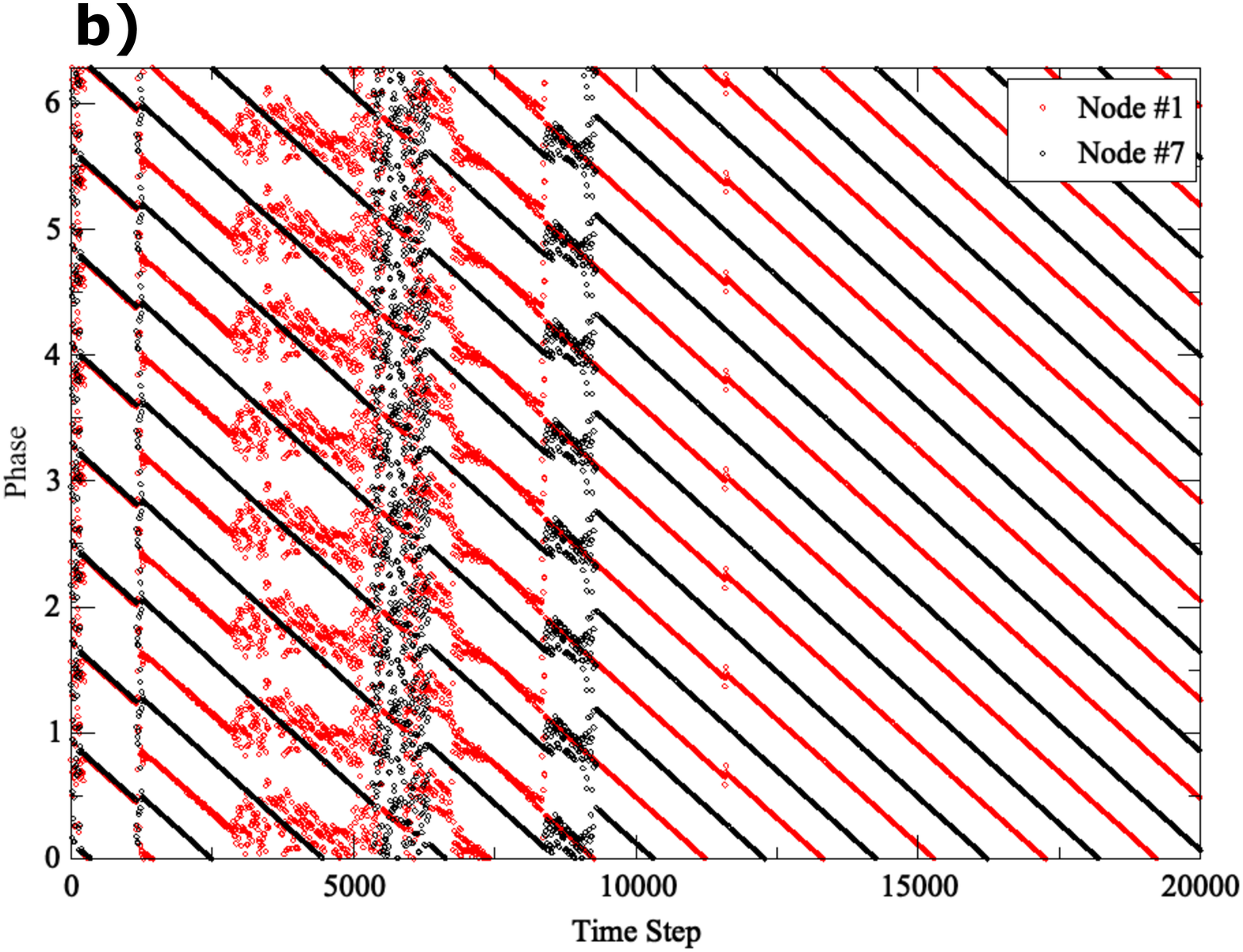}
\end{center}
\caption{An example of computer simulation results.}  
\label{fig:scanner}
\end{figure}

\section{Conclusion}

We can draw the following conclusions:
\begin{enumerate}
\item
More than $\lceil M / N \rceil$ synchronization groups can be non-uniformly distributed in the phase space.
Here, $\lceil x \rceil$ represents the ceiling function of $x$.
\item
Each group is separated from the other by a distance of larger than $\phi_{th}$.
\end{enumerate}

There should be no collisions of selected channels in each group; however, there are some collisions in our results. 
The groups are reorganized if the collisions occur frequently, as shown in Fig. 2(b).   
In this first study, we used TOW dynamics with parameter $\gamma = P_{[1]} + P_{[2]}$ for the channel selection, where  $P_{[n]}$ is the top $n$-th reward probability.
This means that all nodes are selfishly searching for the highest channel only. 
Here, it is known that TOW dynamics can be optimized when $\omega$ $=$ $\gamma $ $/$ ($2$ $-$ $\gamma$).
The optimization in each group will be improved if we use the model setting with $\gamma = P_{[N^{\prime}]} + P_{[N^{\prime}+1]}$ (see ref.~\cite{tow2}), where $N^{\prime}$ represents the number of nodes in the group.

The maximum resource can be estimated as $\frac{2 \pi}{\phi_{th}}$$\times$$\sum_{k=1}^{N}$$P_k$.
In the parameter settings of Fig.2 case, the maximum $12$ ($8$ $\times$ $1.5$) nodes can successfully transmit a packet at every time step.
Our simulation results showed the average $2.86$ nodes, which is about $1/4$ of the maximum resource. 
This is because we used only $M=10$. 
We can use more nodes until $M=40$ (independent $5$ channels $\times$ $8$ phases $=$ $40$ nodes).
Details of performance evaluation for the throughputs of this method will be presented in our future works.

\section*{Acknowledgments}

This work was supported by the research grant SP002 from SOBIN Institute, LLC. 
We would like to thank Dr. Takahito Mitsui (Free University of Berlin) for fruitful discussions in this study.

\urlstyle{tt}

\end{document}